\documentclass[copyright]{eptcs}
\usepackage{breakurl} 

\providecommand{\event}{PLACES 2009} 
\providecommand{\volume}{1}
\providecommand{\anno}{2009}
\providecommand{\firstpage}{1}
\providecommand{\eid}{1}

\usepackage{fancyvrb}
\usepackage{color}

\newcommand{\ttt}[1]{\texttt{#1}}

\title{Programming Idioms for Transactional Events}

\author{Matthew Kehrt \quad Laura Effinger-Dean \quad Michael Schmitz \quad Dan Grossman
\institute{University of Washington \\ Seattle, WA}
\email{\{mkehrt,effinger,schmmd,djg\}@cs.washington.edu}
}

\def\titlerunning{Programming Idioms for Transactional Events}
\def\authorrunning{Kehrt, Effinger-Dean, Schmitz, and Grossman}

\begin{document}
\maketitle

%------------------------------------------------------------------------------
% Abstract
%
\begin{abstract}
  Transactional events (TE) are an extension of Concurrent ML (CML), a
  programming model for synchronous message-passing. Prior work has
  focused on TE's formal semantics and its implementation.  This paper
  considers programming idioms, particularly those that vary
  unexpectedly from the corresponding CML idioms.  First, we solve a
  subtle problem with client-server protocols in TE\@. Second, we
  argue that CML's \ttt{wrap} and \ttt{guard} primitives do not
  translate well to TE, and we suggest useful workarounds.  Finally,
  we discuss how to rewrite CML protocols that use abort actions.
\end{abstract}
%%%%%%%%%%%%% INTRO
\section{Introduction}
\label{sec:intro}
Transactional events (TE) provide powerful message-passing facilities
for concurrent programming \cite{te-hs}.  TE extends Concurrent ML
(CML) \cite{cml-book} with a sequencing primitive \ttt{thenEvt}, which
allows an arbitrary number of sends or receives per event.
\ttt{thenEvt} is powerful enough that programmers can implement patterns
such as $n$-way rendezvous (synchronizing multiple threads at once) or
guarded receive (successfully synchronizing only if a received value
satisfies a predicate).

In previous work \cite{te-hs, te-ml}, we and other researchers
explored the semantics and implementation of transactional events.
However, no existing work discusses practical programming idioms for
TE.  We have found in the course of our research that some each common CML
idioms are surprisingly difficult to reproduce in TE\@.  This paper
presents our solutions to these problematic idioms, which include
client-server protocols and programs using CML's \ttt{wrap},
\ttt{guard}, and \ttt{wrapAbort}.

TE is implemented for Haskell and Caml. In our examples, we use Caml,
which is call-by-value.
%%%%%%%%%%%%% BG
\begin{SaveVerbatim}{verb:types}
type 'a chan
type 'a event
val newChan : unit -> 'a chan
val sync : 'a event -> 'a
val sendEvt : 'a chan -> 'a -> unit event
val recvEvt : 'a chan -> 'a event
val chooseEvt : 'a event -> 'a event
                -> 'a event
val thenEvt : 'a event -> ('a -> 'b event)
              -> 'b event
\end{SaveVerbatim}
\begin{SaveVerbatim}{verb:types2}
val alwaysEvt : 'a -> 'a event
val neverEvt : 'a event
val guard : (unit -> 'a event)
            -> 'a event
val wrap : 'a event -> ('a -> 'b)
           -> 'b event
val wrapAbort : 'a event
                -> (unit -> unit)
                -> 'a event
\end{SaveVerbatim}
\section{Background}
\label{sec:bg}
\begin{figure}[t]
  \fbox{
    \begin{minipage}{.54\textwidth}
      {\BUseVerbatim{verb:types}}
    \end{minipage}
    \begin{minipage}{.42\textwidth}
      {\BUseVerbatim{verb:types2}}
    \end{minipage}
  }
  \caption{Types for several key CML/TE functions.}
  \label{fig:types}
\end{figure}
TE \cite{te-hs,te-ml} and CML \cite{cml-book} are closely related
paradigms for concurrent programming that have been implemented for
several languages.  This section briefly reviews TE and explains how
it differs from CML\@.

In CML, threads send values on channels.  A channel of type
\ttt{'a chan} carries values of type \ttt{'a}.  Communication is
\emph{synchronous}: a send blocks until matched with a receive in
another thread.

An {\em event} is a description of a communication to be
performed. The functions \ttt{sendEvt} and \ttt{recvEvt} produce
events that describe sends and receives, respectively.
\emph{Synchronizing} on an event with the function \ttt{sync} performs
the event; \ttt{sync} has type \ttt{'a event -> 'a}.  Events may be
composed of other events. For example, the function \ttt{chooseEvt}
constructs an event that, when synchronized on, performs exactly one
of two events.  The types of these and other functions appear in
Figure \ref{fig:types}.

TE extends CML with a new function \ttt{thenEvt}, which allows the
sequencing of events.  Synchronizing on \ttt{thenEvt ev f} does the
following: (1) synchronize on \ttt{ev} to produce a result \ttt{v};
(2) apply \ttt{f} to \ttt{v} to produce a new event \ttt{ev2}; (3)
synchronize on \ttt{ev2} to produce a final result.  Critically, these
three steps are all-or-nothing: if the second event cannot
successfully synchronize, then the first event does not (appear to)
happen.  Therefore, we say that events built using \ttt{thenEvt} are
\emph{transactional}.  A single event may communicate with multiple
threads, so the success of one synchronization may entail the success
of an arbitrary number of synchronizations in other threads.

Two more CML/TE functions are useful in combination with
\ttt{chooseEvt} and \ttt{thenEvt}. \ttt{alwaysEvt} is the event that
always succeeds: \ttt{sync (alwaysEvt x)} is equivalent to \ttt{x}.
\ttt{neverEvt} is the event that never succeeds: \ttt{sync neverEvt}
blocks forever.

The following example tries to do either a single send, or a receive
followed by a send:%
\begin{verbatim}
sync (chooseEvt (sendEvt c1 5)
                (thenEvt (recvEvt c2) (fun x -> sendEvt c3 x)))
\end{verbatim}
%%%%%%%%%%%%% SERVER
\section{Server loops}
\label{sec:server}
A common concurrent programming idiom is a server thread that repeatedly handles requests from multiple clients.  In CML, a server is often implemented as an infinite loop that calls \ttt{sync} at every iteration.  In this section, we discuss why TE requires more sophisticated servers and how to implement them.

Consider the following function, which spawns a new thread to act as a server.  The server sends increasing integers on a channel so that clients get a unique integer every time they receive on the channel. 
\begin{verbatim}
let simpleIncrementServer () =
  let c = newChan () in
  let rec f y = sync (sendEvt c y); f (y + 1) in
  Thread.create f 0; c
\end{verbatim}
A simple client receives on the server's channel.
\begin{verbatim}
let simpleIncrementClient c = sync (recvEvt c)
\end{verbatim}
Both client and server are perfectly valid as both TE and CML.  However, in TE \ttt{thenEvt} can be used to write other clients that interact with this server in unexpected ways.  The following code receives two integers and adds them together in a single event:
\begin{verbatim}
let complexIncrementClient c =
  sync (thenEvt (recvEvt c) (fun x ->
        thenEvt (recvEvt c) (fun y ->
        alwaysEvt (x + y))))
\end{verbatim}
If this client and the server were to synchronize on their events, neither would succeed.  The client could receive one integer from the server.  The server event would block waiting for the client event to complete, as they would be participating in the same transaction. Meanwhile, the client would block waiting for another integer.  Both sides would be stuck, so this particular client cannot successfully synchronize with the simple increment server.

A similar problem arises when {\em two} client threads receive an integer from the server and communicate with each other in the same event. A transaction consisting only of these two events and the server event cannot succeed.  The server would need to send to both threads, but the server's event does one send.

To solve this problem, we need a server that can perform multiple sends in a single  call to \ttt{sync}.  In the TE code below, the server synchronizes on one event that can do an arbitrary number of sends.  After each send, the event chooses between returning the sent value or recursively calling the server function.
\begin{verbatim}
let betterIncrementServer () =
  let c = newChan () in
  let rec evtLoop x =
    thenEvt (sendEvt c x)
            (fun _ -> chooseEvt (alwaysEvt (x + 1)) (evtLoop (x + 1))) in
  let rec serverLoop x = serverLoop (sync (evtLoop x)) in
  Thread.create serverLoop 0; c
\end{verbatim}
However, this problem can occur with many different server and client combinations.  A better solution would be to create a generic combinator for an event that is repeated one or more times.

For this purpose, we define a function, \ttt{serverLoop}, suitable for
creating servers.  It takes two arguments and loops forever.  The
first argument is a pair, \ttt{(ev, b)}, of an event and any value.
The second argument is a function, \ttt{f}.  After synchronizing on
\ttt{ev} to produce a value \ttt{a}, \ttt{serverLoop} calls \ttt{f} on
the pair \ttt{(a,b)} to produce a new \ttt{(ev,b)} pair, with which it
recurs.  \ttt{b} acts as a loop-carried state for
\ttt{serverLoop}. Programmers can use \ttt{serverLoop} much like they
use \ttt{fold} to process lists.

Internally, \ttt{serverLoop} uses a second function, \ttt{evtLoop}, that creates an event that synchronizes on \ttt{ev} and then nondeterministically chooses between returning or recurring with the result of calling \ttt{f} to produce a new event and loop-carried state.  In other words, \ttt{evtLoop} does exactly what \ttt{serverLoop} does but, crucially, in a single transaction.
Overall, \ttt{serverLoop} sequentially synchronizes on the events
computed by \ttt{f}, but transactions may end (starting the next transaction)
at any point in the sequence. Thus, clients can communicate with the
server any number of times in one synchronization.
\begin{verbatim}
(* evtLoop :
     ('a event * 'b) -> (('a * 'b) -> ('a event * 'b)) -> ('a * 'b) event *)
let rec evtLoop (ev, b) f =
  thenEvt ev (fun a -> chooseEvt (alwaysEvt (a, b)) (evtLoop (f (a, b)) f))

(* serverLoop : ('a event * 'b) -> (('a * 'b) -> ('a event * 'b)) -> 'c *)
let rec serverLoop (ev, b) f = serverLoop (f (sync (evtLoop (ev, b) f))) f
\end{verbatim}
Using \ttt{serverLoop} to construct an increment server is straightforward.
We spawn a new thread to run \ttt{serverLoop} called with (1)
an initial event paired with the initial loop-carried counter and (2)
a function to construct the next event and counter by incrementing the
counter and creating the next send event.
\begin{verbatim}
let incrementServer () =
  let c = newChan () in
  Thread.create (serverLoop ((sendEvt c 0), 0))
                            (fun (_, x) -> (sendEvt c (x+1), x+1)); c
\end{verbatim}
%%%%%%%%%%%%% WRAP
\section{\texttt{wrap} and \texttt{guard}}
\label{sec:wrap}
\ttt{wrap} and \ttt{guard} (see Figure \ref{fig:types}) add post- and
pre-processing, respectively, to CML events.  \ttt{sync (wrap ev f)}
synchronizes on \ttt{ev}, then applies \ttt{f} to the result.
\ttt{sync (guard g)} synchronizes on the result of \ttt{g ()}.  In
this section, we discuss how to adapt programs that use these
functions to TE.

The following code uses \ttt{wrap} to perform two receives in either
order:
\begin{verbatim}
sync (chooseEvt (wrap (recvEvt c1) (fun x -> (x, sync (recvEvt c2))))
                (wrap (recvEvt c2) (fun x -> (sync (recvEvt c1), x))))
\end{verbatim}
Suppose we were to rewrite this code using \ttt{thenEvt}:
\begin{verbatim}
sync (chooseEvt
  (thenEvt (recvEvt c1)
           (fun x -> thenEvt (recvEvt c2) (fun y -> alwaysEvt (x,y))))
  (thenEvt (recvEvt c2)
           (fun y -> thenEvt (recvEvt c1) (fun x -> alwaysEvt (x,y)))))
\end{verbatim}
These two versions actually behave differently: the CML version
performs the receives in separate synchronizations, while the TE
version executes both in the \emph{same} synchronization.  Therefore
the above TE code could not communicate successfully with code that
performed two synchronizations, such as \ttt{sync (sendEvt c1 4); sync
  (sendEvt c2 5)}.
We can mimic \ttt{wrap}'s behavior in TE by thunking the second
receive and executing the thunk after the first \ttt{sync} completes:
\begin{verbatim}
(sync (chooseEvt
  (thenEvt (recvEvt c1)
           (fun x -> alwaysEvt (fun () -> (x, sync (recvEvt c2)))))
  (thenEvt (recvEvt c2)
           (fun x -> alwaysEvt (fun () -> (sync (recvEvt c1), x)))))) ()
\end{verbatim}
With the use of two helper functions, the TE code approaches the
elegance of the original CML code:
\begin{verbatim}
(* thunkWrap : 'a event -> ('a -> 'b event) -> (unit -> 'b event) *)
let thunkWrap ev f = thenEvt ev (fun x -> alwaysEvt (fun () -> f x))

(* syncThunked : (unit -> 'a) event -> 'a *)
let syncThunked ev = (sync ev) ()

let _ = syncThunked
  (chooseEvt (thunkWrap (recvEvt c1) (fun x -> (x, sync (recvEvt c2))))
             (thunkWrap (recvEvt c2) (fun x -> (sync (recvEvt c1), x))))
\end{verbatim}
We have sacrificed some composability: \ttt{thunkWrap} returns a
\ttt{(unit -> 'b) event} instead of a \ttt{'b event}, so it is more
difficult than \ttt{wrap} to combine with other events.  The semantics
of \ttt{wrap} (processing an event's result after synchronization
completes) and \ttt{thenEvt} (combining two synchronizations into one)
seem to be incompatible, but we believe that thunked wrappers are a
good compromise.  Wrapping an already wrapped event does not require a
second level of thunk, as this helper function demonstrates:
\begin{verbatim}
(* rewrap : (unit -> 'a) event -> ('a -> 'b) -> (unit -> 'b) event *)
let rewrap ev g =
  thenEvt ev (fun f -> let x = f () in alwaysEvt (fun () -> g x))
\end{verbatim}
CML's \ttt{guard} is useful for encapsulating actions that need to
happen prior to synchronization.  For example, the following code adds
a timeout to an event by spawning a thread to signal when to give up:
\begin{verbatim}
(* timeoutEvt : 'a event -> float -> 'a event *)
let timeoutEvt ev time = guard (fun () ->
  let timeoutChan = newChan () in
  Thread.create
    (fun () -> Thread.delay time; sync (sendEvt timeoutChan ())) ();
  chooseEvt ev (wrap (recvEvt timeoutChan) (fun () -> raise TimedOutExn)))
\end{verbatim}

As with \ttt{wrap}, it is difficult to add \ttt{guard} to TE's
interface because the guard function needs to execute outside of the
synchronization.  However, we can code up \ttt{timeoutEvt} in TE
without \ttt{guard}:
\begin{verbatim}
let timeoutEvt ev time = chooseEvt ev
  (thenEvt (alwaysEvt ()) (fun () -> Thread.delay time; raise TimedOutExn))
\end{verbatim}
Moreover, this function is more readable than the CML code. In the
next section, we will see another example for which the TE
implementation is more natural than the original CML program.
%%%%%%%%%%%%% ABORT
\section{\texttt{wrapAbort} and abort actions}
The \ttt{wrapAbort} function (see Figure \ref{fig:types}) lets CML
programs specify an action to take if an event is part of a
\ttt{chooseEvt} that is synchronized on and another choice is taken.
For example, \ttt{sync (chooseEvt (wrapAbort ev1 f) ev2)} will block
until either \ttt{ev1} or \ttt{ev2} succeeds, and in the latter case
it will execute \ttt{f ()}.  \ttt{wrapAbort} is useful for
client-server protocols that have more than one communication. The CML
book \cite{cml-book} uses \ttt{wrapAbort} to implement
mutual-exclusion locks with this interface\footnote{It actually uses
  the equally expressive \ttt{withNack}; we discuss \ttt{wrapAbort}
  because we find it more intuitive.}:
\begin{verbatim}
type lockServer
val acquireLockEvt : lockServer -> int -> unit event
val releaseLockEvt : lockServer -> int -> unit event
val mkLockServer : unit -> lockServer
\end{verbatim}

Clients of this interface acquire or release locks (represented by
\ttt{int}s) by synchronizing on events created with
\ttt{acquire\-Lock\-Evt} and \ttt{releaseLockEvt}. Synchronizing on
\ttt{acquireLockEvt s i} blocks until the acquire succeeds; clients
may abort acquires, perhaps with the \ttt{timeoutEvt} function from
Section \ref{sec:wrap}:
\begin{verbatim}
sync (timeoutEvt (acquireLockEvt server 1) 5.0)
\end{verbatim}

Implementing \ttt{acquireLockEvt} with a single server thread in CML
requires two communications: a request from the client with the lock
ID, and a confirmation from the server when the lock is available.
Abort actions are essential for implementing \ttt{acquireLockEvt}, as
the first communication may affect the server's internal state by
adding the request to a queue. If the client aborts after the first
communication but before the second, it must tell the server to remove
the request from the queue.  Otherwise, the server would hang when
trying to confirm the lock acquire.  The form of \ttt{acquireLockEvt}
is essentially:
\begin{verbatim}
let acquireLockEvt s id = 
  guard (fun () -> (* send acquire request *);
                   wrapAbort (* receive acquire confirmation *)
                             (fun () -> (* do cancellation *)))
\end{verbatim}

The abort action effectively makes the two communications
\emph{transactional}: if the second communication aborts, the effects
of the first communication are canceled.  In TE, we can implement a
lock server without an abort action.  Our solution uses \ttt{thenEvt}
and \ttt{neverEvt} and is similar to the guarded-receive
pattern~\cite{te-hs}.  If the server receives a request for an
unavailable lock (the server maintains a list of held locks), it
returns \ttt{neverEvt} inside \ttt{thenEvt}.  \ttt{neverEvt} never
succeeds, so the program behaves as if the request did not yet occur,
exploiting the transactional semantics of \ttt{thenEvt}.  A full
implementation is below\footnote{As an orthogonal issue, we use
  \ttt{serverLoop} from Section 3 to support multiple lock operations
  in one synchronization.}; we expect other CML protocols that use
abort actions will adapt similarly to TE\@.

\begin{Verbatim}[commandchars=\\\{\}]
type request = Acquire of int | Release of int
type lockServer = request chan
let acquireLockEvt reqCh id = sendEvt reqCh (Acquire id)
let releaseLockEvt reqCh id = sendEvt reqCh (Release id)
let mkLockServer () =
  let reqCh = newChan () in
  let serverEvt heldLocks =
    \textcolor{blue}{thenEvt} (recvEvt reqCh) (function
      | Acquire id -> if List.exists (fun id2 -> id = id2) heldLocks
                          then \textcolor{blue}{neverEvt}
                          else alwaysEvt (id::heldLocks)
      | Release id -> alwaysEvt (List.filter (fun id2 -> id <> id2) heldLocks))
  in Thread.create (serverLoop (serverEvt [], ()))
       (fun (heldLocks, ()) -> (serverEvt heldLocks, ()));
     reqCh
\end{Verbatim}
%%%%%%%%%%%%% CONCLUSION
\section{Conclusion}
\label{sec:conc}
Every programming model needs three things: a semantics, an
implementation, and useful idioms. Reppy's dissertation on CML
\cite{cml-book} presents all three to demonstrate that CML
is a useful programming model.  Prior TE research \cite{te-hs,te-ml}
has concentrated on semantics and implementation. We have presented
several important TE programming idioms, the subtleties of which
surprised us during our research.  First, writing client-server
protocols in TE requires careful consideration of how \ttt{sync}
interacts with \ttt{thenEvt}; our \ttt{serverLoop} function is a
general solution to this problem.  Second, \ttt{wrap} and \ttt{guard}
are difficult to integrate with TE, and we have suggested alternatives
that preserve most of the original semantics.  Finally, we have
discussed how protocols with abort actions may be rewritten
with \ttt{thenEvt} and \ttt{neverEvt}.

%%%%%%%%%%%%% BIBLIOGRAPHY
\bibliographystyle{eptcs}

\end{document}